\documentclass[
    reprint,
    preprintnumbers,
    amsmath,amssymb,
    aps,
    prd
]{revtex4-2}

\usepackage{graphicx}
\usepackage{bm}
\usepackage[dvipsnames]{xcolor}
\usepackage[colorlinks, allcolors=OliveGreen]{hyperref}
\usepackage{array,booktabs,multirow}
\usepackage{enumitem}
\usepackage{mathtools}  

\newcommand{\defeq}{\vcentcolon=}
\newcommand{\defeqrev}{=\vcentcolon}

\newcommand{\secref}[1]{\S\ref{#1}}

\DeclareMathOperator{\tr}{tr}

\renewcommand{\O}{\operatorname{O}}
\DeclareMathOperator{\SU}{SU}
\DeclareMathOperator{\SO}{SO}
\DeclareMathOperator{\Sp}{Sp}

\newcommand{\R}{\mathbb{R}}
\newcommand{\Z}{\mathbb{Z}}
\newcommand{\F}{\mathcal{F}}
\newcommand{\calR}{\mathcal{R}}
\renewcommand{\H}{\mathcal{H}}

\newcommand{\adj}{\mathrm{Adj}}
\newcommand{\rep}[2][]{\mathbf{\underline{#2}^{#1}}}    
\newcommand{\repss}[2]{\mathbf{\underline{#1}_{#2}}}    

\newcommand{\ch}{\mathrm{ch}}

\usepackage{letltxmacro}
\makeatletter
\let\oldr@@t\r@@t
\def\r@@t#1#2{
    \setbox0=\hbox{$\oldr@@t#1{#2\,}$}\dimen0=\ht0
    \advance\dimen0-0.2\ht0
    \setbox2=\hbox{\vrule height\ht0 depth -\dimen0}
    {\box0\lower0.4pt\box2}
}
\LetLtxMacro{\oldsqrt}{\sqrt}
\renewcommand*{\sqrt}[2][\ ]{\oldsqrt[#1]{#2}}
\makeatother

\begin{document}


\preprint{KEK-TH-2600}

\title{
    A new infinite class of \texorpdfstring{$\bm{6d}$, $\bm{\mathcal{N}=(1,0)$}}{6d, N=(1,0)} supergravities
}

\author{Gregory J.\ Loges}
\email{gloges@post.kek.jp}
\affiliation{
    Theory Center, IPNS, High Energy Accelerator Research Organization (KEK),\\
    1-1 Oho, Tsukuba, Ibaraki 305-0801, Japan
}

\date{\today}

\begin{abstract}
    We present a new infinite class of non-abelian, 6d supergravities with eight supercharges.
    These theories not only satisfy all known low-energy consistency conditions, such as being free of anomalies, but also evade the constraints arising from the consistency of string probes, even after assuming BPS completeness.
    This demonstrates that some additional UV input or hitherto unknown IR condition is needed in order to be left with a finite landscape, as is generally anticipated from a theory of quantum gravity.
\end{abstract}

\maketitle


\section{Introduction}
\label{sec:introduction}

In recent years there has been much progress in understanding to what extend string universality holds in higher dimensions. Together, supersymmetry and anomaly cancellation constrain the space of allowed low-energy theories significantly and one can show that there is an exact match with the string theory landscape for $d=8,9,10$~\cite{Adams:2010zy,Kim:2019vuc,Montero:2020icj,Cvetic:2020kuw,Hamada:2021yxy,Hamada:2021bbz}. For $d=7$ there is also a match, up to a classification of three-dimensional $\mathcal{N}=4$ SCFTs~\cite{Bedroya:2021fbu}.

The situation changes dramatically in six dimensions, the maximal dimension which allows for only eight supercharges. While it has been shown that the number of anomaly-free theories is finite provided the number of tensor multiplets is bounded as $T<9$~\cite{Kumar:2010ru}, there \emph{do} appear infinite families for $T\geq 9$. Nevertheless, assuming BPS completeness and using the anomaly-inflow arguments developed in~\cite{Kim:2019vuc} for string probes has proven very effective at truncating all of the (previously) known infinite families to a finite subset~\cite{Kim:2019vuc,Tarazi:2021duw}.

Recently, in~\cite{Hamada:2023zol} it was shown that there is a wide class of infinite families which satisfy all (known) low-energy consistency conditions. These are built starting from a ``seed'' theory, which is very loosely constrained, and augmenting by a huge number of exceptional gauge factors. Since the resulting gauge groups are enormous, it was anticipated that this class of theories would similarly be restricted to a finite number upon considering the constraints imposed by string probes; the purpose of this short article is to show that this is not the case.

The remainder of this article is organized as follows. In \secref{sec:consistency-conditions} we recall the consistency conditions required of 6d supergravities with minimal supersymmetry. Next, in \secref{sec:infinite-families} we present a new class of infinite families of anomaly-free theories in which the gauge group and numbers of tensor- and hypermultiplets are all controlled by a single parameter $m\in\Z_{>0}$ and in \secref{sec:proof} we prove that all of the constraints coming from the consistency of string probes derived in~\cite{Kim:2019vuc} are satisfied for infinitely many values of $m$. Finally, we conclude in \secref{sec:discussion}.


\section{Consistency conditions}
\label{sec:consistency-conditions}

For our purposes, specifying a theory amounts to choosing the following data: the number of tensor multiplets $T\geq 0$, the non-abelian gauge group $G=\prod_i G_i$ with $V=\dim G$ corresponding vector multiplets transforming in the adjoint representation, the total (generally reducible) representation $\H$ of $G$ for $H=\dim\H$ hypermultiplets, and the anomaly vectors $b_I\in\R^{1,T}$ ($I=0,i$)~\footnote{For uniformity in notation we write $b_0 \defeq -a$.}. There is one self-dual 2-form field from the gravity supermultiplet and $T$ anti-self-dual 2-form fields, one from each tensor multiplet, and the vectors $b_I$ control their couplings to the graviton and gauge vectors.

These data cannot be chosen freely. From low-energy considerations alone, gauge and gravity anomalies must be fully cancelled and what we call \emph{positivity} and \emph{unimodularity} conditions must be satisfied.
Assuming BPS completeness (e.g.\ see~\cite{Polchinski:2003bq,Banks:2010zn,Harlow:2018tng}), there must be BPS strings charged under the (anti-)self-dual 2-form fields. When these string probes cannot be coupled to gravity, this signals the theory should be discarded as inconsistent.
We recall each of these conditions in turn.

\subsection{Anomaly cancellation}
\label{sec:anomaly-cancellation}

Gauge and gravitational anomalies may be cancelled by means of the Green-Schwarz-West-Sagnotti mechanism~\cite{Green:1984bx} (see also~\cite{Green:1984sg,Bianchi:1990tb,Bianchi:1990yu,Sagnotti:1992qw,Ferrara:1997gh}) wherein the usual 1-loop contributions from chiral fermions are balanced against the tree-level exchange of the (anti-)self dual 2-form fields. All together, we have
\begin{equation}
    \begin{aligned}
        \hat{I}_8 &= \hat{I}_8^\text{1-L} + \frac{1}{2}Y_4\cdot Y_4 \,,\\
        Y_4^\alpha &\defeq -\frac{1}{2}b_0^\alpha \tr\calR^2 + 2\sum_i \frac{b_i^\alpha}{\lambda_i}\tr\F_i^2 \,,
    \end{aligned}
\end{equation}
where $Y_4\in\R^{1,T}$ is a vector of 4-forms and the $\lambda_i$ are normalization constants given in Table~\ref{tab:lambda-coxeter}. The theory is free of local anomalies exactly when $\hat{I}_8=0$. For each irreducible term in $\hat{I}_8$ there is a corresponding constraint:
\begin{align}
    \label{eq:gravitational-constraint}
    \tr\mathcal{R}^4 &:&                      H-V+29T &= 273 \,, \qquad\;\; \\
    \label{eq:B-constraint}
    \tr\F_i^4        &:& \sum_R n_R^iB_R^i - B_\adj^i &= 0   \,.
\end{align}
The remaining, reducible terms determine all inner products amongst the vectors $b_I$:
\begin{equation}
    \begin{aligned}
        \left(\tr\mathcal{R}^2\right)^2 &:&
            b_0\cdot b_0 &= 9-T \,, \\
        \left(\tr\F_i^2\right)^2 &:&
            b_i\cdot b_i &= \frac{1}{3}\Big(\sum_R n_R^iC_R^i - C_\adj^i\Big)\,, \\
        \tr\calR^2 \tr\F_i^2 &:&
            b_0\cdot b_i &= \frac{1}{6}\Big(\sum_R n_R^iA_R^i - A_\adj^i\Big) \,, \\
        \tr\F_i^2 \tr\F_j^2 &:&
            b_i\cdot b_j &= \sum_{R,S}n_{(R,S)}^{i,j}A_R^iA_S^j \,,\;\; (i\neq j) \,.
    \end{aligned}
\end{equation}
$n_R^i$ gives the number of hypermultiplets in the representation $R$ of the gauge factor $G_i$. Of course, since the vectors $b_I$ live in $\R^{1,T}$, the matrix of inner products $b_I\cdot b_J$ can have at most one positive eigenvalue and at most $T$ negative eigenvalues. These bounds on the signature of $b_I\cdot b_J$ are actually necessary and sufficient to ensure that there exist vectors $b_I\in\R^{1,T}$ which realize the inner products dictated by the massless spectrum. In cases where $b_I\cdot b_J$ has a positive eigenvalue or $T$ negative eigenvalues, the vectors $b_I\in\R^{1,T}$ are uniquely determined by their inner products up to $\O(1,T;\R)$ transformations.

\begin{table}[t]
    \begin{tabular}{>{$\;}c<{\;$}|*{8}{>{$\;}c<{\;$}}}
        \toprule
        G_i          & \SU(N) & \SO(N) & \Sp(N) & E_6 & E_7 & E_8 & F_4 & G_2 \\ \midrule
        \lambda_i    & 1      & 2      & 1      & 6   & 12  & 60  & 6   & 2   \\
        h_i^\lor & N      & N-2    & N+1    & 12  & 18  & 30  & 9   & 4   \\ \bottomrule
    \end{tabular}
    \caption{Normalization constants and dual Coxeter numbers for simple gauge factors.}
    \label{tab:lambda-coxeter}
\end{table}

In relating traces in a representation $R$ of $G_i$ to the trace in the fundamental, we have introduced the indices $A_R^i$, $B_R^i$ and $C_R^i$, defined through
\begin{equation}
\label{eq:ABC-defn}
    \begin{aligned}
        \lambda_i\tr_R\F_i^2 &= A_R^i\tr\F_i^2 \,, \\
        \lambda_i^2\tr_R\F_i^4 &= B_R^i\tr\F_i^4 + C_R^i\left(\tr\F_i^2\right)^2 \,.
    \end{aligned}
\end{equation}
With this choice of normalization $A_R^i$, $B_R^i$ and $C_R^i$ are nearly always integers~\footnote{The only exceptions are for $A_1\sim\SU(2)$, $A_2\sim\SU(3)$, $B_3\sim\SO(7)$ and $D_4\sim\SO(8)$, where $C_R^i\in\frac{1}{2}\Z$.} and we have $A_\adj^i=2h_i^\lor$ with $h_i^\lor$ the duel Coxeter number of $G_i$. See Table~\ref{tab:irrep-indices} for the indices of some common irreducible representations with this normalization. For the simple groups $\SU(2)$, $\SU(3)$, $E_n$, $F_4$ and $G_2$ there is no independent quartic Casimir invariant and thus $B_R^i=0$ for all representations $R$. Finally, there are additional potential global anomalies for $\SU(2)$, $\SU(3)$ and $G_2$ gauge factors~\cite{Bershadsky:1997sb,Ohmori:2014kda,Lee:2020ewl,Davighi:2020kok} (manifesting as constraints modulo $12$, $6$ and $3$, respectively) which are easily avoided. It has been shown that the absence of global anomalies follows from the absence of local anomalies and that all of the inner products $b_I\cdot b_J$ are integers~\cite{Kumar:2010ru}. That is, $\Lambda\defeq\bigoplus_I b_I\Z \subset \R^{1,T}$ is an integral lattice and the eigenvalue bounds discussed above are no longer sufficient.

\begin{table}[t]
    \begin{tabular}{>{$}l<{$}*{4}{>{$\,}r<{\,$}}}
        \toprule
        \multicolumn{1}{c}{$G$} & \multicolumn{1}{c}{$\qquad R$} & A_R & B_R & C_R \\ \midrule
        \multirow{2}{*}{$\SU(2)$}
            & \rep{2}            & 1    & 0     & 1/2 \\
            & \rep{3}            & 4    & 0     & 8  \\[5pt]
            & \rep{3}            & 1    & 0     & 1/2 \\
        \SU(3) & \rep{6}         & 5    & 0     & 17/2 \\
            & \rep{8}            & 6    & 0     & 9  \\[5pt]
        \multirow{4}{*}{$\SU(N\geq 4)\hspace{-20pt}$}
            & \rep{N}            & 1    & 1     & 0  \\
            & \rep{N(N-1)/2}     & N-2  & N-8   & 3  \\
            & \rep{N(N+1)/2}     & N+2  & N+8   & 3  \\
            & \rep{N^2-1}        & 2N   & 2N    & 6  \\[5pt]
        \multirow{4}{*}{$\SO(N)$}
            & \rep{N}            & 2    & 4     & 0  \\
            & \rep{(N-1)(N+2)/2} & 2N+4 & 4N+32 & 12 \\
            & \rep{2^{\lfloor\frac{N-1}{2}\rfloor}}
                & 2^{\lfloor\frac{N-5}{2}\rfloor}
                & -2^{\lfloor\frac{N-5}{2}\rfloor}
                & 3\cdot2^{\lfloor\frac{N-9}{2}\rfloor} \\
            & \rep{N(N-1)/2}     & 2N-4 & 4N-32 & 12 \\[5pt]
        \multirow{3}{*}{$\Sp(N)$}  
            & \rep{2N}           & 1    & 1     & 0  \\
            & \rep{(N-1)(2N+1)}  & 2N-2 & 2N-8  & 3  \\
            & \rep{N(2N+1)}      & 2N+2 & 2N+8  & 3  \\[5pt]
        \multirow{2}{*}{$E_6$}
            & \rep{27}           & 6    & 0     & 3  \\
            & \rep{78}           & 24   & 0     & 18 \\[5pt]
        \multirow{2}{*}{$E_7$}
            & \rep{56}           & 12   & 0     & 6  \\
            & \rep{133}          & 36   & 0     & 24 \\[5pt]
        E_8 & \rep{248}          & 60   & 0     & 36 \\[5pt]
        \multirow{2}{*}{$F_4$}
            & \rep{26}           & 6    & 0     & 3  \\
            & \rep{52}           & 18   & 0     & 15 \\[5pt]
        \multirow{2}{*}{$G_2$}
            & \rep{7}            & 2    & 0     & 1  \\
            & \rep{14}           & 8    & 0     & 10 \\
        \bottomrule
    \end{tabular}
    \caption{Common irreducible representations and their indices $A_R$, $B_R$, $C_R$ defined by Eqn.~\eqref{eq:ABC-defn}.}
    \label{tab:irrep-indices}
\end{table}

\subsection{Positivity and unimodularity}
\label{sec:positivity-unimodularity}

Scalars in the tensor multiplets parametrize the tensor branch of the moduli space, $j\in \R^{1,T}$ with $j\cdot j=1$. The gauge kinetic terms are $-j\cdot b_i\tr\F_i^2$, so we require that there exists some choice for $j\in\R^{1,T}$ with $j\cdot j>0$ that gives $j\cdot b_i>0$ for all $i$. The analogous quantity $j\cdot b_0$ controls the coefficient of the Gauss-Bonnet term, and although we do not demand that it is strictly positive we will see that in fact $j\cdot b_0>0$ holds for the class of theories constructed in~\secref{sec:infinite-families}.

As we saw above, the absence of anomalies implies that the lattice $\Lambda$ is integral. $\Lambda$ is a sub-lattice of the string charge lattice $\Gamma$ and in~\cite{Seiberg:2011dr} it was shown by reducing to four and two dimensions that $\Gamma$ must be a unimodular (i.e.\ integral and self-dual) lattice of signature $(1,T)$. For the examples discussed in \secref{sec:infinite-families}, this condition will be manifestly satisfied since we will realize $b_I$ directly as elements of the odd unimodular lattice $\Z^{1,T}$. We note that in general, even if the lattice $\Lambda\subset\R^{1,T}$ is completely fixed up to $\O(1,T;\R)$ transformations, there may be several inequivalent ways to realize $\Lambda$ as a sub-lattice of $\Gamma$ (i.e.\ with different $\Gamma/\Lambda$ or equivalently no $\O(1,T;\Z)$ transformation relating them).

\subsection{String probes and anomaly inflow}
\label{sec:string-probes}

\begin{figure}[t]
    \includegraphics[width=\linewidth]{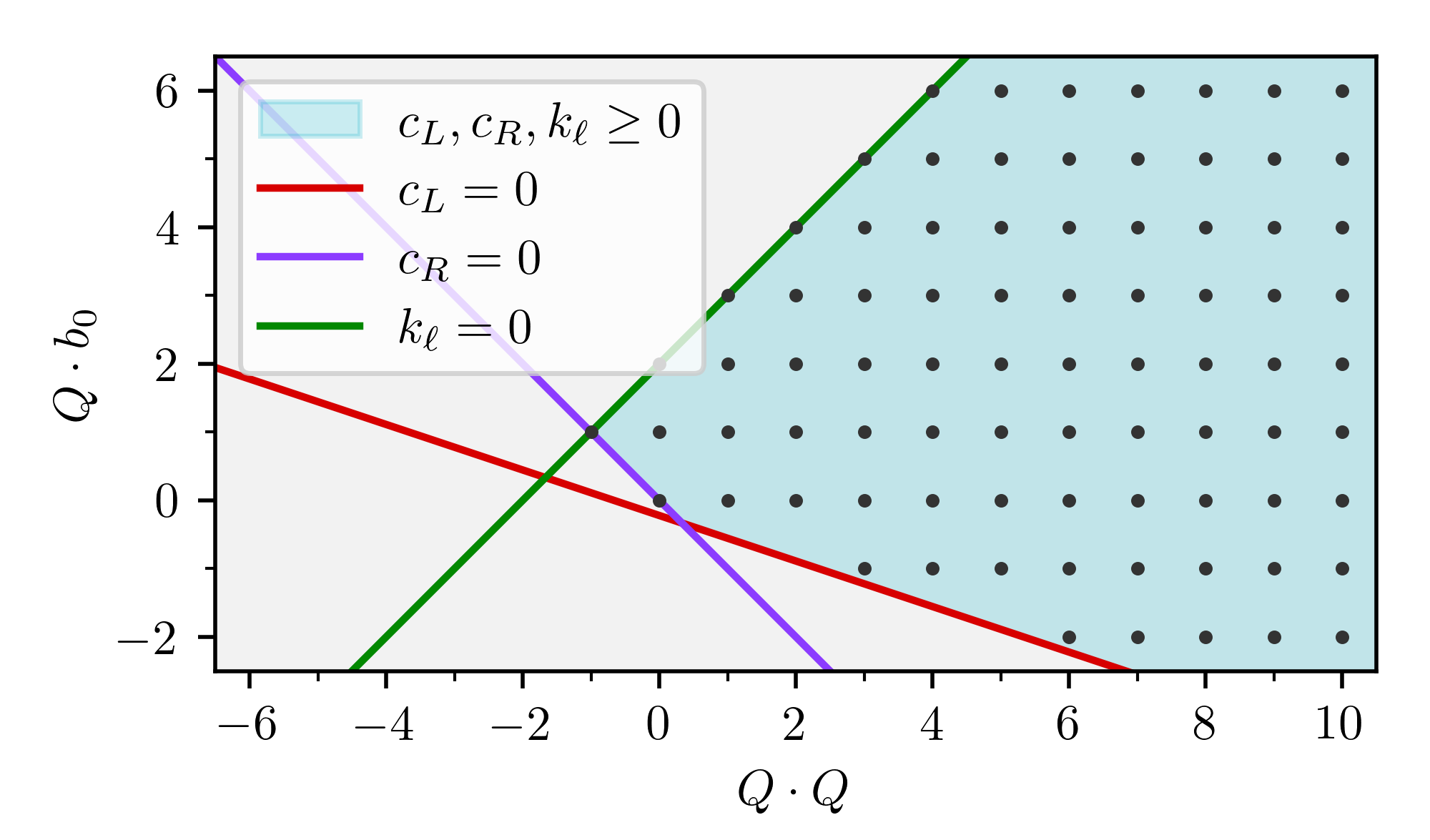}
    \caption{Values for $Q\cdot Q$ and $Q\cdot b_0$ which are allowed by the bounds $c_L\geq 0$, $c_R\geq 0$ and $k_\ell\geq 0$.}
    \label{fig:QQ-Qb0-values}
\end{figure}

When a string probe charged under the 2-form fields is introduced into a background configuration, in general this induces anomalies on the worldsheet which can then be cancelled by the anomaly-inflow mechanism~\cite{Callan:1984sa,Blum:1993yd,Freed:1998tg,Shimizu:2016lbw,Kim:2019vuc}.
The central charges were computed in~\cite{Kim:2019vuc}: after subtracting off the center-of-mass contributions which decouple in the IR, these read
\begin{equation}
    \begin{aligned}
        c_L &= 3Q\cdot Q + 9Q\cdot b_0 + 2 \,, \\
        c_R &= 3Q\cdot Q + 3Q\cdot b_0 \,, \\
        k_\ell &= \tfrac{1}{2}\big(Q\cdot Q - Q\cdot b_0 + 2\big) \,, \\
        k_i &= Q\cdot b_i \,.
    \end{aligned}
\end{equation}
$c_L$ and $c_R$ are the gravitational central charges for the left- and right-moving sectors and the levels $k_\ell$ and $k_i$ correspond to $\SU(2)_\ell$ and the bulk gauge symmetry $G_i$, respectively. When all quantities are non-negative,
\begin{equation}
\label{eq:admissible-defn}
    c_L,\,c_R,\,k_\ell,\,k_i \geq 0 \,,
\end{equation}
a non-trivial constraint arises from requiring that $c_L$ is large enough to accommodate a unitary representation of the current algebra, namely
\begin{equation}
\label{eq:BPS-inequality}
    \sum_i c_i \leq c_L \,, \qquad c_i \defeq \frac{k_i\dim G_i}{k_i + h_i^\lor} \,.
\end{equation}
Showing that a theory is inconsistent amounts to an existential statement:
\begin{equation}
    \exists\, Q\in\Gamma \;:\; c_L,c_R,k_\ell,k_i\geq 0 \;\;\text{and}\;\; \sum_ic_i > c_L \,.
\end{equation}
In contrast, showing that a theory is \emph{not} revealed to be inconsistent via anomaly inflow is a universal statement and can be considerably more difficult to establish:
\begin{equation}
\label{eq:forall-Q}
    \forall\, Q\in\Gamma \,,\;\; c_L,c_R,k_\ell,k_i\geq 0 \;\implies\; \sum_ic_i \leq c_L \,.
\end{equation}
The computations of~\secref{sec:proof} amount to a proof of exactly this statement for the class of theories we will describe. There, it will be convenient to refer to charges $Q\in\Gamma$ satisfying Eqn.~\eqref{eq:admissible-defn} as \emph{admissible}. The region of the $Q\cdot Q$ vs.\ $Q\cdot b_0$ plane carved out by three of the conditions, $c_L\geq 0$, $c_R\geq 0$ and $k_\ell\geq 0$, is shown in Fig.~\ref{fig:QQ-Qb0-values}; we will make repeated use of the following weaker bounds,
\begin{align}
    c_R,k_\ell &\geq 0 & &\implies& Q\cdot Q &\geq -1               \,,\label{eq:QQ-bound}\\
        k_\ell &\geq 0 & &\implies&      c_L &\geq 12Q\cdot b_0 - 4 \,,\label{eq:cL-bound}
\end{align}
the second clearly only being useful for $Q\cdot b_0\geq 1$.

\section{A new class of infinite families}
\label{sec:infinite-families}

\subsection{Preliminaries}

Our starting point will be a ``seed'' infinite family with simple, non-abelian group $G_\text{seed}$ and charged hypermultiplets $\H_\text{seed}^\ch$ chosen so that
\begin{equation}
\label{eq:seed-shape}
    2\delta \defeq b^\text{seed}\cdot b^\text{seed} - b_0\cdot b^\text{seed}
\end{equation}
is constant. This requires $\H_\text{seed}^\ch$ to be of the form
\begin{equation}
\label{eq:H0+xH1}
    \H_\text{seed}^\ch = \H_0^\ch \oplus x\H_1^\ch \,,
\end{equation}
where we take both $\H_0^\ch$ and $\H_1^\ch$ to be some fixed (non-trivial) representations of $G_\text{seed}$. The hypermultiplets in $\H_0^\ch$ determine $\delta$ and must satisfy the constraint of Eqn.~\eqref{eq:B-constraint}, while the hypermultiplets in $\H_1^\ch$ must satisfy
\begin{align}
\label{eq:H1-requirements}
    \sum_Rn_R^iB_R^i &= 0 \,, & \sum_R n_R^iA_R^i &= 2\sum_R n_R^iC_R^i
\end{align}
so that $x\geq 0$ can be chosen freely. For each such seed, we can construct the following infinite family depending only on $m\geq 1$,
\begin{equation}
\label{eq:infinite-family}
    \begin{aligned}
        G &= G_\text{seed}\times E_8^{4m} \,,\\
        \H^\ch &= (\H_\text{seed}^\ch, \rep{1}^{4m}) \,,\\
        T &= 12m+1 \,,
    \end{aligned}
\end{equation}
where we choose $x$ to grow with $m$ in such a way that
\begin{equation}
\label{eq:gamma-C-defn}
    \begin{aligned}
        b^\text{seed}\cdot b^\text{seed} \!&\;= 24m + 4C_\delta\gamma_\delta \\
        \gamma_\delta &\defeq \begin{cases}
            \frac{\delta+2}{2} & \delta\equiv 0\mod{2} \,,\\
            \frac{\delta+3}{4} & \delta\equiv 1\mod{4} \,,\\
            \frac{\delta-3}{4} & \delta\equiv 3\mod{4} \,,
        \end{cases}\\
        C_\delta &\defeq \begin{cases}
            2 & \delta \equiv 0\mod{2} \,,\\
            3 & \delta \equiv 1\mod{2} \,.
        \end{cases}
    \end{aligned}
\end{equation}
The (integer) constants $\gamma_\delta$ and $C_\delta$ have been introduced for later convenience: the reason for having different cases based on the value of $\delta$ modulo four is to allow for $b_I\in\Z^{1,T}$, as we will see shortly. Let us immediately check that the gravitational anomaly can be cancelled for arbitrarily large $m$. Using Eqn.~\eqref{eq:H0+xH1}, if we separate the contributions from $\H_0^\ch$ and $\H_1^\ch$ and write
\begin{equation}
    \begin{aligned}
        H_\text{seed}^\ch &= H_0^\ch + xH_1^\ch \,,\\
        b^\text{seed}\cdot b^\text{seed} &= (b\cdot b)_0 + x(b\cdot b)_1 \,,
    \end{aligned}
\end{equation}
then from Eqn.~\eqref{eq:gamma-C-defn} we have $x=(\text{const.}) + \frac{24m}{(b\cdot b)_1}$ and
\begin{align}
    H^\ch - V + 29T &= (\text{const.}) - 24m\big(\tfrac{161}{6} - \tfrac{H_1}{(b\cdot b)_1}\big) \,.
\end{align}
Therefore
\begin{equation}
\label{eq:H1-bound}
    \frac{H_1^\ch}{(b\cdot b)_1} < \frac{161}{6} = 26.8333\ldots
\end{equation}
is required so that $H^\ch-V+29T$ decreases with $m$. By taking $m$ large enough we can ensure that $H^\ch-V+29T\leq 273$ for \emph{any} choice of $\H_0^\ch$ and then by adding in an appropriate number of neutral hypermultiplets, the constraint of Eqn.~\eqref{eq:gravitational-constraint} can be met exactly. However, the simple group $G_\text{seed}$ is indirectly restricted to have small rank via Eqn.~\eqref{eq:H1-bound}: a complete list of possibilities for $G_\text{seed}$ and $\H_1^\ch$ is given in Table~\ref{tab:Gseed-H1-examples}.

From Eqns.~\eqref{eq:seed-shape},~\eqref{eq:infinite-family} and~\eqref{eq:gamma-C-defn}, the inner products are
\begin{align}
\label{eq:Gram}
    b_I\cdot b_J &= \begin{psmallmatrix}
        8-12m & 24m+4C_\delta\gamma_\delta-2\delta & -10 \\
        24m+4C_\delta\gamma_\delta-2\delta & 24m+4C_\delta\gamma_\delta & 0\\
        -10 & 0 & \!\!\!-12\,\mathbb{I}_{4m\times 4m}
    \end{psmallmatrix} \,.
\end{align}
These can be realized by the integer vectors
\begin{equation}
\label{eq:bI-vectors}
    \begin{aligned}
        b_0 &= \big(3;\,1,\,1^{12m}\big) \,,\\
        b^\text{seed} &=  \big(W_\delta(m) + C_\delta;\,-W_\delta(m) + C_\delta,\, (-\vec{u}_\delta)^{m}\big) \,,\\
        b_{4r+t}^{E_8} &= \big({-1};\,1,\,0^{12r},\, \vec{v}_t,\, 0^{12(m-r-1)} \big) \,,
    \end{aligned}
\end{equation}
where $r\in\{0,1,\ldots,m-1\}$ and $t\in\{1,2,3,4\}$. We have introduced both
\begin{align}
    W_\delta(m) &\defeq (6-C_\delta)m + \gamma_\delta
\end{align}
and the following 12-component vectors:
\begin{equation}
\label{eq:vt-vectors}
    \begin{aligned}
        \vec{u}_\delta &\defeq \begin{cases}
            (0^4,1^8) \,, & \delta\equiv 0\mod{2} \,,\\
            (1^{12})  \,, & \delta\equiv 1\mod{2} \,,
        \end{cases}\\
        \vec{v}_1 &\defeq (-1,\phantom{-}1,\phantom{-}1,\phantom{-}1,\;2,\;2,\;0,\;0,\;0,\;0,\;0,\;0) \,,\\
        \vec{v}_2 &\defeq (\phantom{-}1,-1,\phantom{-}1,\phantom{-}1,\;0,\;0,\;2,\;2,\;0,\;0,\;0,\;0) \,,\\
        \vec{v}_3 &\defeq (\phantom{-}1,\phantom{-}1,-1,\phantom{-}1,\;0,\;0,\;0,\;0,\;2,\;2,\;0,\;0) \,,\\
        \vec{v}_4 &\defeq (\phantom{-}1,\phantom{-}1,\phantom{-}1,-1,\;0,\;0,\;0,\;0,\;0,\;0,\;2,\;2) \,.
    \end{aligned}
\end{equation}
Also, it is straightforward to check that
\begin{equation}
\label{eq:j-example}
    j \;\propto\; b^\text{seed} - \epsilon\cdot\sum_{i=1}^{4m} b_i^{E_8}
\end{equation}
gives time-like $j$ with $j\cdot b_0>0$, $j\cdot b^\text{seed}>0$ and $j\cdot b_i^{E_8}>0$ when $24m+4C_\delta\gamma_\delta >\max\{0, 2\delta\}$ and $\epsilon>0$ is small enough.

\begin{table}[t]
    \centering
    \begin{tabular}{>{$\;}l<{\;$}>{$\quad}l<{\quad$}>{$\;}c<{\;$}>{$\;}c<{\;$}}
        \toprule
        \multicolumn{1}{c}{$G_\text{seed}\;\;$} & \multicolumn{1}{c}{$\H_1^\ch$} & H_1^\ch & (b\cdot b)_1\\
        \midrule

        \SU(2)  & 6\times\rep{2}                 & 12 & 1 \\
        \SU(3)  & 6\times\rep{3}                 & 18 & 1 \\
        \SU(4)  & 4\times\rep{4} \oplus \rep{6}  & 22 & 1 \\
        \SU(5)  & 3\times\rep{5} \oplus \rep{10} & 25 & 1 \\[5pt]

        \SO(7)  & \hspace{0.58em} \rep{7} \oplus 2\times\rep{8}  & 23 & 1 \\
        \SO(8)  & \hspace{0.08em} \repss{8}{v} \oplus \repss{8}{s} \oplus \repss{8}{c} & 24 & 1 \\
        \SO(9)  & \hspace{0.58em} \rep{9} \oplus \rep{16}        & 25 & 1 \\
        \SO(10) &                 \rep{10} \oplus \rep{16}       & 26 & 1 \\[5pt]

        \Sp(2)  &      4\times\rep{4} \oplus \rep{5}  & 21 & 1 \\
        \Sp(3)  & \scalebox{0.8}{$\begin{aligned}
            &(2x_1 + \tfrac{7}{2}x_2)\times\rep{6}\\[-3pt]
            &{} \oplus x_1\times\rep{14} \oplus (\tfrac{1}{2}x_2)\times\rep[\prime]{14}
        \end{aligned}$} & 26x_1 + 28x_2 & x_1 + x_2 \\[5pt]

        F_4     & \phantom{3\times{}} \rep{26}       & 26 & 1 \\
        G_2     &                     3\times\rep{7} & 21 & 1 \\

        \bottomrule
    \end{tabular}
    \caption{All simple groups $G_\text{seed}$ and hypermultiplets $\H_1^\ch$ which satisfy Eqns.~\eqref{eq:H1-requirements} and~\eqref{eq:H1-bound}. For $\Sp(3)$ there is more freedom in choosing $\H_1^\ch$, but the non-negative, co-prime integers $x_1$, $x_2$ must satisfy $7x_2 < 5x_1$.}
    \label{tab:Gseed-H1-examples}
\end{table}

\smallskip

In summary, we conclude that all of the anomaly cancellation, positivity and unimodularity conditions are met, provided Eqn.~\eqref{eq:H1-bound} is satisfied and $m$ is taken large enough to ensure that $H^\ch-V+29T\leq 273$ and $24m+4C_\delta\gamma_\delta > \max\{0,2\delta\}$. In the next section we will make use of one additional mild lower bound on $m$,
\begin{align}
    \frac{C_\delta W_\delta(m) - 3m}{C_\delta^2} &> 1 & \Longleftrightarrow & & m > \frac{C_\delta(C_\delta - \gamma_\delta)}{9 + C_\delta} \,,\label{eq:WC-condition}
\end{align}
which is non-trivial only for $\delta=-1,1,3,7$ or $\delta\leq -3$. Also, for the most part we will only need the \emph{average} $E_8$ anomaly-vector,
\begin{equation}
    b_\text{avg}^{E_8} \defeq \frac{1}{4m}\sum_{i=1}^{4m}b_i^{E_8} = \big({-1};\,1,\, (\tfrac{1}{2m})^{12m}\big) \,.
\end{equation}
This must satisfy $k_\text{avg}^{E_8}\defeq Q\cdot b_\text{avg}^{E_8}\geq 0$ as well, but $k_\text{avg}^{E_8}$ is clearly no longer necessarily integer-valued.

\subsection{Constraints from string probes are satisfied}
\label{sec:proof}

It only remains to show that the constraints imposed by string probes are satisfied for all admissible $Q$. While we have in mind the situation where $m$ is large (as must ultimately occur if this is to be an infinite family), this will only serve to guide the analysis and suggest a line of attack: we make no approximations or large-$m$ expansions and all of the inequalities we derive, while not necessarily sharp, are exact.

By design, the vectors $b_I$ have the following three key features: (i) there are no ``free'' components of $b_0$ which could be leveraged to decrease $Q\cdot Q$ and $Q\cdot b_0$ without altering $Q\cdot b_i$, (ii) the components of $b^\text{seed}$ and each quartet $\sum_{t=1}^4 b_{4r+t}^{E_8}$ are all of opposite sign (or zero), and (iii) since 
\begin{equation}
    \begin{aligned}
        \vec{v}_1 + \vec{v}_2 + \vec{v}_3 + \vec{v}_4 = (2,2,\ldots,2) \,,
    \end{aligned}
\end{equation}
\emph{on average} the contributions to $Q\cdot b_i^{E_8}$ from each of these groups of 12 components are proportional to the corresponding contributions to $Q\cdot b_0$. This is the reason why the number of $E_8$ factors was chosen to be a multiple of four. These three features together lead to two mechanisms which will ensure that $c_L$ on the right-hand side of Eqn.~\eqref{eq:BPS-inequality} outpaces the left-hand side as $m$ increases. The first is that the cone described by $Q\cdot b^\text{seed}\geq 0$ and $Q\cdot b_i^{E_8}\geq 0$ is restricted to be quite narrow thanks to (ii), and due to charge quantization the smallest non-zero admissible $Q$ is therefore necessarily large. The second is that together (i) and (iii) will ultimately force $Q\cdot b_0$ to be positive which, as we saw in Eqn.~\eqref{eq:cL-bound}, also provides a non-trivial lower bound $c_L\geq 12Q\cdot b_0-4$.

\medskip

We now set out to prove that~\eqref{eq:forall-Q} holds for all large-enough $m$. To begin, write the string charge as
\begin{equation}
\label{eq:Q-vector}
    Q \defeq \left(\tfrac{1}{2}(q_+ + q_-);\,\tfrac{1}{2}(q_+ - q_-),\, q_1,\,\ldots,\,q_{12m}\right) \,,
\end{equation}
where the ``light-cone charges'' $q_\pm$ must have the same parity. The inner products with $Q$ are
\begin{align}
    Q\cdot Q                  &= q_+ q_- - \sum(q_a)^2 \,,\label{eq:Q.Q}\\
    Q\cdot b_0                &= q_+ + 2q_- - \sum q_a \,,\label{eq:Q.b0}\\
    Q\cdot b^\text{seed}      &= W_\delta(m)q_+ + C_\delta q_- + \sum\nolimits' q_a \,,\label{eq:Q.bseed}\\
    Q\cdot b_\text{avg}^{E_8} &= -q_+ - \frac{1}{2m}\sum q_a \,.\label{eq:Q.bavg}
\end{align}
where in order to reduce clutter we will leave all sums over $a\in\{1,2,\ldots,12m\}$ unadorned, other than primes such as in $Q\cdot b^\text{seed}$ above which indicate that for $\delta$ even the indices $a\equiv 1,2,3,4\mod{12}$ are omitted: cf.~Eqn.~\eqref{eq:vt-vectors}. Notice that if $G_\text{seed}$ and the corresponding vector $b^\text{seed}$ were absent, then the charge
\begin{align}
    Q = (0;\, -1,\, 0^{12m}) \,,
\end{align}
for which $c_L=8$, $c_R=0$, $k_\ell=0$ and $k_i^{E_8}=1$, would immediately reveal the theories with $G=E_8^{4m}$ and no charged hypermultiplets to be inconsistent for \emph{all} $m\geq 1$ \footnote{Note that with our ansatz the $E_8$ factors must come in groups of four. Theories with gauge group $E_8$ and $E_8\times E_8$ are perfectly consistent~\cite{Seiberg:1996vs,Kim:2019vuc}.}. However, the requirement of having non-negative level, $Q\cdot b^\text{seed}\geq 0$, presents an obstacle to choosing such a charge.

\subsubsection{An upper bound}

To warm up, let us first bound the left-hand side of Eqn.~\eqref{eq:BPS-inequality} from above. Using the inequality
\begin{equation}
    \frac{1}{N}\sum_{n=1}^N \frac{x_n}{x_n+1} \leq \frac{\sum_{n=1}^N x_n}{\sum_{n=1}^N(x_n + 1)} \,,
\end{equation}
which holds for any set of $N$ non-negative real numbers~\footnote{This is the AM-HM inequality applied to the numbers $\frac{1}{x_n+1}$ in disguise.}, we have
\begin{equation}
    \sum_{i=1}^{4m} \frac{Q\cdot b_i^{E_8}}{Q\cdot b_i^{E_8} + 30} \leq 4m\times\frac{Q\cdot b_\text{avg}^{E_8}}{Q\cdot b_\text{avg}^{E_8} + 30} \,.
\end{equation}
Therefore
\begin{align}
\label{eq:upper-bound}
    \sum_i c_i &\leq \frac{992\,k_\text{avg}^{E_8}}{k_\text{avg}^{E_8}+30}\,m + \dim G_\text{seed} \,,\\
    &\leq \min\!\Big\{ 1,\, \tfrac{k_\text{avg}^{E_8}}{30}\Big\}\times 992m + \dim G_\text{seed} \,. \notag
\end{align}
It will be important that this quantity grows as quickly as $992m$ only if $k_\text{avg}^{E_8}\gtrsim 30$ is appreciable. For small $k_\text{avg}^{E_8}$ the growth with $m$ is greatly reduced.

\subsubsection{A bound on \texorpdfstring{$q_\pm$}{q-pm}}

Next, we claim that $q_\pm > 0$ is required for a non-zero charge to be admissible. The alternatives are each quickly ruled out in turn:
\begin{itemize}[leftmargin=*]
    \item $(q_+q_-<0)$: From $Q\cdot Q\geq -1$ we clearly need $q_a=0$ and $|q_\pm|=1$. However, using $W_\delta(m)>C_\delta$ (which follows easily from Eqn.~\eqref{eq:WC-condition}) in~\eqref{eq:Q.bseed} and~\eqref{eq:Q.bavg} we see that no choice of signs for $q_\pm$ allows for both $Q\cdot b^\text{seed}\geq 0$ and $Q\cdot b_\text{avg}^{E_8}\geq 0$.
    
    \item $(q_+q_-=0)$: From $Q\cdot Q\geq -1$ we learn that at most one of the $q_a$ is nonzero, say $q_b$, and that $|q_b|\leq 1$. Then from $Q\cdot b^\text{seed}\geq 0$ and $Q\cdot b_\text{avg}^{E_8}\geq 0$ the only possibility is $q_+=0$ and $q_-\geq 0$. However, $k_\ell\geq 0$ now gives
    \begin{equation}
        2(1-q_-) \geq 2(1-q_-) + q_b(1-q_b) \geq 0 \,,
    \end{equation}
    and since $q_-\equiv q_+\mod{2}$ we must have $q_-=0$ as well. If $q_b=0$, then $Q=0$ and we are done. Otherwise, if $Q\cdot b^\text{seed} = q_b$ we must have $q_b=1$ but then $Q\cdot b_0=-1$ and $c_L\geq 0$ is violated. For $\delta$ even, it is possible to have $Q\cdot b^\text{seed}=0$ independent of $q_b=\pm1$ when $b\equiv 1,2,3,4\mod{12}$, but then from the form of $\vec{v}_t$ it is clear that we cannot choose $q_b$ so that all $Q\cdot b_i^{E_8}$ are non-negative.
    
    \item $(q_\pm < 0)$: From $Q\cdot b^\text{seed}\geq 0$ we find
    \begin{equation}
        \Big(\sum\nolimits' q_a\Big)^2 \geq \big[W_\delta(m)q_+ + C_\delta q_-\big]^2 \,,
    \end{equation}
    and so
    \begin{equation}
        \begin{aligned}
            Q\cdot Q &\leq q_+q_- - \sum\nolimits'(q_a)^2 \,,\\
            &\leq q_+q_- - \frac{1}{12m}\Big(\sum\nolimits' q_a\Big)^2 \,,\\
            &\leq q_+q_- - \tfrac{1}{12m}\big[W_\delta(m)q_+ + C_\delta q_-\big]^2 \,,\\
            &= -\tfrac{C_\delta W_\delta(m) - 3m}{C_\delta^2}(q_+)^2 \\
            &\qquad - \tfrac{1}{12m}\left(\tfrac{(C_\delta W_\delta(m)-6m)q_+ + C_\delta^2 q_-}{C_\delta}\right)^2 \,,\\
            &\leq -\tfrac{C_\delta W_\delta(m) - 3m}{C_\delta^2}(q_+)^2 \,,
        \end{aligned}
    \end{equation}
    violating $Q\cdot Q\geq -1$ thanks to Eqn.~\eqref{eq:WC-condition}. The second line above follows from the Cauchy-Schwarz inequality.
\end{itemize}

\subsubsection{A lower bound on \texorpdfstring{$c_L$}{cL}}

Continuing with $q_\pm > 0$, we immediately have that $Q\cdot b_0$ is positive by combining Eqns.~\eqref{eq:Q.b0} and~\eqref{eq:Q.bavg}:
\begin{equation}
\label{eq:Q.b0-substituted}
    Q\cdot b_0 = (2m+1)q_+ + 2q_- + 2m\,k_\text{avg}^{E_8} \,.
\end{equation}
If we use the crude bounds $q_\pm \geq 1$ then we find $Q\cdot b_0\geq 2m(1 + k_\text{avg}^{E_8}) + 3$ and $c_L\geq 24m(1 + k_\text{avg}^{E_8}) + 32$ after using Eqn.~\eqref{eq:cL-bound}. However, this is not sufficient to ensure that Eqn.~\eqref{eq:BPS-inequality} is satisfied since it is well below the upper bound from Eqn.~\eqref{eq:upper-bound}. By finding better bounds on $q_\pm$ we will be able to improve the lower bound on $c_L$.

Inspired by Fig.~\ref{fig:QQ-Qb0-values}, we should expect that the most constraining bound comes from $k_\ell\geq 0$ since we have already established that $Q\cdot b_0\geq 1$. Using Eqns.~\eqref{eq:Q.Q},~\eqref{eq:Q.b0} and~\eqref{eq:Q.bavg} in $k_\ell\geq 0$, we find
\begin{equation}
    \newcommand{\ww}{\hspace{30pt}}
    \begin{aligned}
        &(q_+ - 2)(q_- - 1) \\
        &\ww\geq \sum(q_a)^2 - \sum q_a \,,\\
        &\ww\geq \frac{1}{12m}\Big(\sum q_a\Big)\Big(\sum q_a - 12m\Big) \,,\\
        &\ww= \frac{m}{3}\big(q_+ + k_\text{avg}^{E_8}\big)\big(6 + q_+ + k_\text{avg}^{E_8}\big) \,,
    \end{aligned}
\end{equation}
again using Cauchy-Schwarz for the second inequality. For large $m$ we must have $q_-\gtrsim\mathcal{O}(mq_+)$ which means that $Q$ roughly aligns with $b^\text{seed}$ and thus also $j$ in Eqn.~\eqref{eq:j-example}, as expected: note that together $b^\text{seed}\cdot b^\text{seed}>0$, $Q\cdot Q>0$, $j\cdot j>0$, $Q\cdot b^\text{seed}>0$ and $j\cdot b^\text{seed}>0$ imply that the string's tension $j\cdot Q$ is automatically positive. The right-hand side above is manifestly positive and therefore we must have $q_+\geq 3$ and $q_-\geq 2$. Already this improves the earlier bounds to $Q\cdot b_0\geq 2m(3 + k_\text{avg}^{E_8}) + 7$ and $c_L\geq 24m(3 + k_\text{avg}^{E_8}) + 80$, but still this is insufficient. We can clearly do much better since the right-hand side above is ${\geq\!9m}$; dividing through by $q_+ - 2>0$ to bound $q_-$ and using Eqn.~\eqref{eq:Q.b0-substituted}, we find
\begin{align}
    c_L &\geq 12Q\cdot b_0 - 4 \,,\notag\\
    &= 24m\big(q_+ + k_\text{avg}^{E_8}\big) + (12q_+ + 20) + 24(q_- - 1) \,,\notag\\
    &\geq 24m\big(q_+ + k_\text{avg}^{E_8}\big) + 56 \notag\\
    &\qquad + \frac{8m(q_+ + k_\text{avg}^{E_8})(6 + q_+ + k_\text{avg}^{E_8})}{q_+-2} \,,\\
    &= \frac{8(q_+ + k_\text{avg}^{E_8})(4q_+ + k_\text{avg}^{E_8})}{q_+-2}\,m + 56  \,,\notag\\
    &\defeqrev C(q_+,k_\text{avg}^{E_8})\,m + 56 \,. \notag
\end{align}
A quick calculation shows that for fixed $k_\text{avg}^{E_8}\geq 0$ and $q_+\geq 3$, $C(q_+,k_\text{avg}^{E_8})$ is minimized at
\begin{equation}
    q_+^\ast \defeq 2 + \frac{1}{2}\sqrt{\big(k_\text{avg}^{E_8} + 2\big)\big(k_\text{avg}^{E_8} + 8\big)} \geq 4 \,,
\end{equation}
and that
\begin{equation}
    C(q_+^\ast,k_\text{avg}^{E_8}) \geq 256 + 72\,k_\text{avg}^{E_8} \,.
\end{equation}
This gives us our final lower bound on $c_L$:
\begin{equation}
\label{eq:lower-bound}
    c_L \geq \big(256 + 72\,k_\text{avg}^{E_8}\big)m + 56 \,.
\end{equation}

\subsubsection{Summary}

In summary, we have shown that for all non-zero charges $Q\in\Gamma$ satisfying Eqn.~\eqref{eq:admissible-defn},
\begin{align}
    c_L &\geq \big(256 + 72\,k_\text{avg}^{E_8}\big)m + 56 \,,\\
    \sum_i c_i &\leq \frac{992\,k_\text{avg}^{E_8}}{k_\text{avg}^{E_8} + 30}\,m + \dim G_\text{seed} \,,
\end{align}
both hold for large-enough $m$. It is readily checked that
\begin{equation}
\label{eq:m-coefficient-ineq}
    256 + 72\,k_\text{avg}^{E_8} \gg \frac{992\,k_\text{avg}^{E_8}}{k_\text{avg}^{E_8}+30}
\end{equation}
for all $k_\text{avg}^{E_8}\geq 0$ so that, given that $\dim G_\text{seed}$ is fixed, we can always ensure Eqn.~\eqref{eq:BPS-inequality} is satisfied by taking $m$ large enough. Therefore we conclude that, provided only that Eqn.~\eqref{eq:H1-bound} is met, the theories of Eqn.~\eqref{eq:infinite-family} satisfy all of the consistency conditions, including those stemming from the consistency of string probes, for infinitely many values of $m$.


\section{Discussion}
\label{sec:discussion}

In this article we have demonstrated that the landscape of consistent 6d supergravities with eight supercharges and non-abelian gauge group is infinite, even after assuming BPS completeness and requiring the consistency of all string probes. This is in stark contrast to the situation in higher dimensions where supersymmetry and anomalies together lead to an exact match with the finite string landscape.

The examples we have constructed have very few objectionable features other than their large number of degrees of freedom and inclusion of many exceptional groups: all of the hypermultiplets can be chosen to be in standard representations (e.g.\ fundamental, two-index (anti-)symmetric, spinor and adjoint); we have the usual choice $b_0=(3;1^T)$ which is both a characteristic and primitive vector of the string charge lattice; although non-zero admissible charges must be large, there is no unnatural hierarchy since $Q\cdot Q \gtrsim\mathcal{O}(m)$ arises from requiring $Q$ have non-negative inner product with $\mathcal{O}(m)$ distinct vectors of norm $\mathcal{O}(1)$ and one vector of norm $\mathcal{O}(m)$.

For concreteness we have considered theories where $H^\ch$, $V$ and $T$ are tied together through the single parameter $m$. Given that not all of the inequalities are sharp and Eqn.~\eqref{eq:m-coefficient-ineq} is satisfied by such a wide margin, it seems reasonable to expect that this is just the tip of the iceberg and more general examples of a similar nature could be found. There are a few obvious places for generalization:
\begin{enumerate}
    \item The auxiliary $E_8$ gauge factors used here can likely be replaced by $E_6$, $E_7$ or $E_7+\tfrac{1}{2}\rep{56}$ (with the vectors $b_I$ adjusted accordingly) without too much trouble: these are the four combinations identified in~\cite{Hamada:2023zol} as leading to infinite families with $T$ unbounded.
    
    \item We took $b^\text{seed}\cdot b^\text{seed}$ and $b_0\cdot b^\text{seed}$ to grow at the same rate with $m$ (cf.\ Eqn.~\eqref{eq:seed-shape}); one could imagine relaxing this and allowing for different constants of proportionality in place of the ``$2$'' in Eqn.~\eqref{eq:H1-requirements}, which likely would change the requirement of Eqn.~\eqref{eq:H1-bound}.
    
    \item We considered cases where $G_\text{seed}$ is a simple group, but it is possible to have $G_\text{seed}$ semi-simple. As a simple example, one can choose
    \begin{align*}
        G_\text{seed} &= \Sp(2)\times \Sp(2) \,,\\
        \H_\text{seed}^\ch &= (\rep{10},\rep{1}) \oplus (\rep{1},\rep{10}) \\
        &\qquad {}\oplus x\big[(\rep{4},\rep{4}) \oplus (\rep{5},\rep{1}) \oplus (\rep{1},\rep{5})\big] \,,\\
        &\hspace{-10pt} b_i^\text{seed} \cdot b_j^\text{seed} = b_0^\text{seed}\cdot b_i^\text{seed} = x \,,
    \end{align*}
    and by taking $b_1^\text{seed}=b_2^\text{seed}$ the analysis of \secref{sec:infinite-families} continues to hold with essentially no changes. It may be possible to find more general families with semi-simple $G_\text{seed}$ and vectors $b_I$ which realize the same mechanisms leveraged here.
\end{enumerate}
A key feature of the class of theories presented here, however, appears to be that $H^\ch$ is unbounded: attempts to adapt Eqn.~\eqref{eq:infinite-family} to have $\H_\text{seed}^\ch$ (and thus also $b^\text{seed}\cdot b^\text{seed}$) constant were unsuccessful, although perhaps an entirely different structure for the vectors $b_I$ which facilitates this could be engineered.

\medskip

How can we recover a finite landscape, as generally expected from a theory of quantum gravity?
Certainly there are no known ways to construct theories with an unbounded number of gauge factors or tensor multiplets from string theory.
The class of theories constructed here provide very strong guidelines for any future attempts to definitively prove finiteness of the supergravity landscape. For example, it is not enough to bound the rank of individual gauge factors or limit the possible hypermultiplet representations since the examples above have $\operatorname{rank} G_i\leq 8$ and hypermultiplets can be chosen to only appear in fundamental representations: the proposals of~\cite{Tarazi:2021duw} are easily met.

There appears to be two possible ways forward. (i) Some universal bound on one of $T$, $V$ or $H^\ch$, perhaps in connection with the species scale~\cite{Dvali:2007hz,Dvali:2009ks,Dvali:2012uq} (which here clearly decreases rapidly with $m$), places an upper bound on $m$. (ii) Additional \emph{global} anomalies kill these families at large $T\sim m$, such as those of Dai-Freed type recently studied in~\cite{Basile:2023zng} for $T\leq 1$ \footnote{My thanks to I.\ Basile and G.\ Leone for bringing this possibility to my attention.}.
It may also be fruitful to consider the introduction of brane probes of other dimensionality, although their presence is then no longer guaranteed by the completeness hypothesis.
We leave demoting this class of infinite families to the swampland for the bright future.

\bigskip


\begin{acknowledgments}
    My thanks to both Yuta Hamada and Gary Shiu for comments on an earlier draft, and to the members and visitors of the Harvard Swampland Initiative for their comments and questions.
    The work of G.L.\ is supported in part by MEXT Leading Initiative for Excellent Young Researchers Grant Number JPMXS0320210099.
\end{acknowledgments}


\bibliography{refs}

\end{document}